\def\fullversionflag{1}     

\ifodd\fullversionflag
  \documentclass[letterpaper,twocolumn,11pt]{article}
  \usepackage[cm]{fullpage}
  \usepackage{times}
\else
  \documentclass[sigconf]{acmart}
\fi

\usepackage{etex}
\usepackage{etoolbox}

\newtoggle{fullversion}
\ifodd\fullversionflag
    \toggletrue{fullversion}
\else
    \togglefalse{fullversion}
\fi

\newcommand{\ignore}[1]{}

\usepackage{amsfonts}
\usepackage{amsmath,amsthm}
\usepackage{latexsym}
\usepackage{graphicx}
\usepackage{subcaption}
\usepackage{xcolor}
\usepackage{pgfplots}
\usepackage{enumitem}
\usepackage{fancybox}
\usepackage{balance} 
\iftoggle{fullversion}{
\usepackage[colorlinks=true,citecolor=purple,urlcolor=violet]{hyperref}
}{}

\usepackage{listings}

\usepackage{listings, xcolor}

\definecolor{verylightgray}{rgb}{.97,.97,.97}

\lstdefinelanguage{Solidity}{
	keywords=[1]{anonymous, assembly, assert, balance, break, call, callcode, case, catch, class, constant, continue, constructor, contract, debugger, default, delegatecall, delete, do, else, emit, event, experimental, export, external, false, finally, for, function, gas, if, implements, import, in, indexed, instanceof, interface, internal, is, length, library, log0, log1, log2, log3, log4, memory, modifier, new, payable, pragma, private, protected, public, pure, push, require, return, returns, revert, selfdestruct, send, solidity, storage, struct, suicide, super, switch, then, this, throw, transfer, true, try, typeof, using, value, view, while, with, addmod, ecrecover, keccak256, mulmod, ripemd160, sha256, sha3}, 
	keywordstyle=[1]\color{blue}\bfseries,
	keywords=[2]{address, bool, byte, bytes, bytes1, bytes2, bytes3, bytes4, bytes5, bytes6, bytes7, bytes8, bytes9, bytes10, bytes11, bytes12, bytes13, bytes14, bytes15, bytes16, bytes17, bytes18, bytes19, bytes20, bytes21, bytes22, bytes23, bytes24, bytes25, bytes26, bytes27, bytes28, bytes29, bytes30, bytes31, bytes32, enum, int, int8, int16, int24, int32, int40, int48, int56, int64, int72, int80, int88, int96, int104, int112, int120, int128, int136, int144, int152, int160, int168, int176, int184, int192, int200, int208, int216, int224, int232, int240, int248, int256, mapping, string, uint, uint8, uint16, uint24, uint32, uint40, uint48, uint56, uint64, uint72, uint80, uint88, uint96, uint104, uint112, uint120, uint128, uint136, uint144, uint152, uint160, uint168, uint176, uint184, uint192, uint200, uint208, uint216, uint224, uint232, uint240, uint248, uint256, var, void, ether, finney, szabo, wei, days, hours, minutes, seconds, weeks, years},	
	keywordstyle=[2]\color{teal}\bfseries,
	keywords=[3]{block, blockhash, coinbase, difficulty, gaslimit, number, timestamp, msg, data, gas, sender, sig, value, now, tx, gasprice, origin},	
	keywordstyle=[3]\color{violet}\bfseries,
	identifierstyle=\color{black},
	sensitive=true,
	comment=[l]{//},
	morecomment=[s]{/*}{*/},
	commentstyle=\color{gray}\ttfamily,
	stringstyle=\color{red}\ttfamily,
	morestring=[b]',
	morestring=[b]"
}

\renewcommand{\paragraph}[1]{\medskip\noindent\textbf{#1}}
\newcommand{\Section}[1]{\hyperref[#1]{Section~\ref*{#1}}}

\newcommand{\deq}{\mathrel{\mathop:}=}
\newlist{pseudocode}{enumerate}{2}
\setlist[pseudocode]{label=\raisebox{0.5px}{\scriptsize{\sf \arabic*:}}, ref=\arabic*, labelwidth=2em, labelsep=1em, align=right, itemsep=.2ex,topsep=0.5ex}

\ignore{
\usepackage{microtype}

}

\title{R-Pool and Settlement Markets for Recoverable ERC-20R Tokens}

\iftoggle{fullversion}{}{
\acmConference{ACM CCS DeFi Workshop}{Nov. 30, 2023}{Copenhagen, Denmark}
}

\iftoggle{fullversion}{
  \usepackage{authblk}
  \author[*]{Kaili Wang}
  \author[**]{Qinchen Wang}
  \author[*]{Calvin Cai}
  \author[**]{Dan Boneh}
  \affil[*]{Circle Inc., New York, NY, USA}
  \affil[**]{Stanford University, Stanford, CA, USA}
}{
  \author{Kaili Wang}
  \affiliation{\institution{Circle Inc.} \city{New York} \state{NY} \country{USA}}
  \email{kaili.wang@circle.com}
  \author{Qinchen Wang}
  \affiliation{\institution{Stanford University} \city{Stanford} \state{CA} \country{USA}}
  \email{qinchenw@cs.stanford.edu}
  \author{Calvin Cai}
  \affiliation{\institution{Circle Inc.} \city{New York} \state{NY} \country{USA}}
  \email{ccai@circle.com}
  \author{Dan Boneh}
  \affiliation{\institution{Stanford University} \city{Stanford} \state{CA} \country{USA}}
  \email{dabo@cs.stanford.edu}
}

\iftoggle{fullversion}{
  \newcommand{\publication}[1] {\noindent\vspace*{-1em}\raisebox{77mm}[0mm][0mm]{\hspace*{-0mm}\noindent\parbox[t]{165mm}{\sl{#1}}}}
}{
  \copyrightyear{2023}
  \acmYear{2023}
  \setcopyright{acmlicensed}\acmConference[DeFi '23]{Proceedings of the 2023 Workshop on Decentralized Finance and Security}{November 30, 2023}{Copenhagen, Denmark}
  \acmBooktitle{Proceedings of the 2023 Workshop on Decentralized Finance and Security (DeFi '23), November 30, 2023, Copenhagen, Denmark.}
  \acmPrice{15.00}
  \acmDOI{10.1145/3605768.3623542}
  \acmISBN{979-8-4007-0261-7/23/11}
}

\begin{document}

\maketitle

\publication{An extended abstract of this paper appears in the 
       Proceedings of the 2023 ACM Workshop on 
       Decentralized Finance and Security, ACM DeFi 2023,
       Nov. 2023, Copenhagen, Denmark}

\begin{abstract} 
ERC-20R is a wrapper around ERC-20 that supports asset recovery within a limited time window after an asset is transferred.  
It is designed to reduce theft and losses on the blockchain 
by allowing a victim to recover their stolen or lost assets during the recovery window.
When an honest recipient receives an ERC-20R asset, 
they must wait until the recovery windows elapses (say, 24 hours),
before they can unwrap the asset back to its base ERC-20 form. 
We argue that many DeFi services will likely refuse to accept unsettled recoverable assets
because they can interfere with their normal operations. 
Consequently, when Alice receives an ERC-20R token, she must wait 24 hours before she can use it with a DeFi service.
But what if Alice is willing to pay a fee to exchange the wrapped token for an unwrapped ERC-20 token that can be used right away?
In this paper we explore how to design a pool to exchange an unsettled ERC-20R asset for a base ERC-20 of the same asset.
Designing such a pool raises several challenging questions and we present our solutions. 
\end{abstract}

\iftoggle{fullversion}{}{


}
  
\section{Introduction} \label{sec:introduction}

Annual losses due to token theft and accidental losses are in the billions:
in 2021, \$3.3 billion was stolen in crypto hacks, and that number jumped to \$3.8 billion in 2022~\cite{chainalysis_2023}.
Funds lost or stolen are often irrecoverable due to the irreversible nature of the blockchain.
One way to protect assets from theft is to strengthen the security of asset keys and to improve the quality of Web3 code.
A complementary approach, explored in depth by Wang, Wang, and Boneh~\cite{ERC20r}, 
is to extend the ERC-20 interface to support asset recovery.

We briefly review the ERC-20R architecture and refer to~\cite{ERC20r} for a detailed description.
Recoverable ERC-20, called ERC-20R, gives the victim of a theft a short time window, say 24 hours, to initiate an adjudicated recovery process.  
ERC-20R is a wrapper contract around an existing ERC-20 contract ---
an ERC-20 token can be wrapped to become an ERC-20R token, as explained in \Section{sec:wrapper}.
When a transaction transfers ERC-20R tokens from Alice's account,
she has 24 hours to issue a freeze request to an arbitration contract, along with evidence of the theft.
If the request is approved, the arbitration contract tracks Alice's funds according to a prescribed algorithm~\cite{ERC20r}, 
starting from the initial theft recipient, and instructs the ERC-20R contract to freeze the stolen assets in the tainted accounts 
(alternatively, the ERC-20R developer can choose to deploy their own asset tracking algorithm). 
This begins a manual arbitration process (e.g., as in~\cite{erc792}) that ends in one of two outcomes:
either the freeze is lifted, or the funds are returned to Alice.  
We give more details in \Section{sec:wrapper}.

An interesting consequence of ERC-20R is that Alice's token balance is split in two: 
settled tokens~$S$ and unsettled tokens~$U$. 
Her balance in the ERC-20R contract is the sum of $S$ and $U$.  
When a transaction deposits funds into Alice's account, the funds are initially deposited into her unsettled balance $U$.
During the freeze window (say, 24 hours) these funds are subject to a freeze, and a possible clawback.
Once the freeze window elapses, the funds can no longer be frozen, 
and they are transferred from Alice's unsettled balance~$U$ to her settled balance~$S$.  
We stress that settled tokens cannot be clawed back from Alice by a recovery process. 
In addition, Alice can unwrap a {\em settled} token from her balance $S$ back into a base ERC-20 token, {\em with no delay},
as discussed in \Section{sec:wrapper}.

Since ERC-20R was introduced, several projects have built on it.
A few examples include 
the \href{https://novo-space.webflow.io/}{\color{violet}Novospace} project,
the \href{https://resolv.finance}{\color{violet}Resolve} project,
as well as some recovery \href{https://zora.co/collect/eth:0x5908eb01497b5d8e53c339ea0186050d487c8d0c/44}{\color{violet}UX proposals}.
In addition, a company called {\sf Lossless.io} has previously deployed a closely related architecture. 

\paragraph{ERC-20R and DeFi.}
DeFi services, such as exchanges and lending protocols, do no expect assets to be clawed back.
For example, suppose that a liquidity provider contributes unsettled tokens to a Uniswap liquidity pool.
It is possible that those tokens simply disappear from the pool due to a recovery event. 
The Uniswap contract, and most DeFi protocols, are not built to handle such an event.  
Consequently, it is likely that DeFi protocols will refuse to accept wrapped ERC-20R tokens, and only accept base ERC-20 tokens.

Nevertheless, when a DeFi service receives ERC-20 tokens, it would be wise to immediately wrap the tokens in an ERC-20R wrapper.
This protects the service's token pool from bugs and hacks that could drain the pool.
In case of a hack, the service operator could request a freeze and begin the recovery process. 
When the DeFi service sends tokens from its pool to another account, it sends them as wrapped ERC-20R tokens.  
If the transfer was caused by a hack, it could get the tokens back through a recovery process. 
In other words, DeFi services that use ERC-20R introduce an asymmetry:  
they only accept base ERC-20 tokens, but send out wrapped ERC-20R tokens. 

Let us examine the implication of this asymmetry. 
When an honest Bob receives a wrapped ERC-20R token, 
he cannot use it right away because DeFi services refuse to accept wrapped tokens.
Bob cannot unwrap the token into a base ERC-20 token because the token is currently unsettled. 
He can use the token to pay someone else, but they would be unable to use it with a DeFi service.
As such, Bob's only option is to wait 24 hours for the token to become settled, unwrap it,
and then use it with the desired DeFi service. 

But what if Bob wants to use the unsettled token as soon as it is received?
He is even willing to pay a fee to unwrap his unsettled token without delay. 
This raises the need for a special purpose pool that we call an {\bf R-Pool}:
\begin{center}
\setlength\shadowsize{1px}
\shadowbox{\quad \parbox{0.8\columnwidth}{
An {\bf R-Pool} is a pool used to exchange {\em unsettled} ERC-20R tokens for base ERC-20 tokens. 
}\quad} \end{center}
The design of such a pool is the main topic of this paper. 

\paragraph{Designing an R-pool.}
We explore two possible designs for an R-pool: 
an automated R-pool and an order book R-pool. 

The automated R-pool interface looks like a regular Uniswap pool interface. 
The pool holds both settled and unsettled ERC-20R tokens (it does not hold base ERC-20 tokens). 
A liquidity provider (LP) can send unwrapped base tokens to the pool, which then get wrapped into settled tokens. 
In return, the LP receives LP tokens that record its stake in the pool.
Bob can use the R-pool to exchange his unsettled tokens for base ERC-20 tokens using a $(1-\epsilon)$ exchange rate.
The pool will keep Bob's unsettled tokens, unwrap some of its settled tokens, and send back the resulting base ERC-20 tokens.
At a later time, the liquidity provider can withdraw their tokens by converting their LP tokens back to a mix of ERC-20R and base tokens. 

The order book version is a simpler and more straightforward design, where users can post orders to the contract and LPs evaluate the risk of clawback themselves before matching an order.

Clearly, an LP who contributes to any R-pool is taking a risk:
unsettled tokens might become frozen in the pool, or even clawed back, and the LP will not be able to get their full deposit back. 
However, LPs understand this risk and are compensated accordingly.

\medskip\noindent
Designing an R-pool raises a number of challenges:
\begin{itemize}[itemsep=2px]
\item 
How do we prevent a thief from using the R-pool to launder its stolen tokens?

\item
What should the unsettled to settled exchange rate be?
The rate will depend on the state of the pool, but also on the clawback risk associated with a specific transaction.
Unsettled ERC-20R tokens are not fungible --- they have different risk profiles --- and this makes it harder to set an exchange rate. 

\item 
What happens when a liquidity provider withdraws their tokens from the pool?
Do they get settled or unsettled tokens?
If the pool suffered a loss from a clawback, how much should the LP get back?
\end{itemize}
In the next few sections we present our answers to these questions. 

An interesting aspect of the R-pool is that the ratio of settled to unsettled tokens 
is constantly changing, even when no one is using the pool. 
For example, 24 hours after Bob sends his unsettled tokens to the pool, 
they automatically become settled and are moved from the unsettled portion of the pool to the settled portion.
If there is no user activity for 24 hours, then the entire unsettled portion of the pool goes to zero.
As such, the constant product formula is not a good fit for setting the exchange rate for an R-pool. Another reason why the constant product formula is not a good fit for setting the exchange rate is that it does a poor job of reflecting the level of risk for clawback; each trade's risk of clawback may depend on the provenance of the unsettled tokens and the user's transaction history. Thus, a different mechanism is needed.

\section{The ERC-20R Wrapper} 
\label{sec:wrapper}

We begin by giving some more details about the ERC-20R architecture.
The architecture is implemented and the code will be made available in an open source repository.

\medskip\noindent
Every ERC-20R contract has two primary preset parameters:
\begin{itemize}[itemsep=1px,topsep=2px,parsep=1px]
\item the address of the arbitration contract that can freeze and recover tokens, and 
\item the time window during which a transferred token is recoverable (e.g. 24 hours).
\end{itemize}

\noindent
ERC-20R works best as an ERC-20 wrapper, rather than a token itself. 
First, it makes recoverability an optional feature to add to existing assets. Wrapping can be thought of as ``protecting'' the asset.
Second, for a given base asset, different parties may need different configurations. 
Some may want a longer or shorter settlement period. Similarly, different parties may want different arbitrators for the same asset. 
Third, wrapping provides a much easier deployment path as there is no need to change the base asset. 

\paragraph{Wrapping and unwrapping.}
When a base ERC-20 asset gets wrapped with ERC-20R, 
a {\em settled} ERC-20R token is minted while the base asset is locked in the ERC-20 contract. 
Going the other way, anyone can unwrap a {\em settled} token with no delay, 
and this transfers a base ERC-20 token to the caller.
An unsettled token cannot be unwrapped until it becomes settled. 

Why is there no delay when unwrapping a settled ERC-20R token?
Recall that settled tokens cannot be clawed back from their owner. 
Hence, the only risk is that the current owner of the tokens has been
compromised,
and the attacker is trying to evade recovery by unwrapping the tokens 
while they are in the owner's possession and then transferring the 
unwrapped tokens to the attacker's address.
If the owner is a DeFi contract, then the attacker would need to 
cause the contract to call the {\em unwrap} function,
and this is not always possible when a bug is being exploited 
or an oracle is being manipulated.
Moreover, since most DeFi contracts do not ever need to unwrap tokens
themselves (they transfer them to others in wrapped form),
there is a {\em disableUnwrap} function that a contract can call 
when it is first created to prevent any future unwraps. 
Another argument for immediate unwrap of settled tokens is the 
far better user experience compared to a 24-hour delayed unwrap.

\paragraph{Transferring ERC20R tokens.}
When Alice sends tokens to Bob, she can choose to send them from either her {\em settled} balance or her {\em unsettled} balance.
Either way the tokens are deposited into Bob's unsettled balance.
Bob can choose to send the just-received unsettled tokens to Carol, then Carol can send them to David, and so on.
If Alice initiates a recovery process within 24 hours of sending the tokens to Bob, 
there is a discovery algorithm~\cite{ERC20r} that locates the funds on chain (say, in David's possession) and freezes them. 

\paragraph{The ERC-20R interface.}
The interface for ERC-20R (IERC-20R) has the same six functions as the IERC-20 interface, 
except that many of these functions take an additional boolean flag as input that indicates whether the action should 
be applied to the settled or unsettled portion of the account's balance.
If no flag is specified then it defaults to false (the settled portion). 
For example, here is the interface for two functions:
\begin{lstlisting}[language=Solidity,basicstyle=\tiny]
function balanceOf(address account, bool includeUnsettled) external view returns (uint256);
function transfer(address to, uint256 amount, bool includeUnsettled) external returns (bool);
\end{lstlisting} 
IERC-20R also supports the following additional functions:
\begin{lstlisting}[language=Solidity,basicstyle=\tiny]
function baseToken() external view returns (address); 
function wrap(uint256 amount) external;
function unwrap(uint256 amount) external;
function unwrapTo(uint256 amount, address to) external;
function nonce(address account) external view returns (uint256);
function disableUnwrap() external;
\end{lstlisting}
The {\sf baseToken} function returns the address of the base ERC-20 contract.
The {\sf unwrapTo} function will remove tokens from the caller's settled ERC-20R balance
and send the same number of base ERC-20 tokens to the specified address.
The {\sf nonce} function returns the current nonce associated with an account, which increments by one every time the account receives or sends ERC-20R funds. 
The purpose of this nonce will be discussed in the next section. 
The {\sf disableUnwrap} function causes all subsequent unwrap requests from the caller's account to fail. Once unwraps for an account are disabled, it cannot be re-enabled. This is an optional security measure for addresses (e.g. contracts) that only send out wrapped tokens and do not need to unwrap their ERC-20R tokens themselves.

By definition, the ERC-20R interface also supports a {\tt freeze} function and a {\tt recover} function. 
We do not list them as part of the interface because they are only 
callable by the arbitration contract and are 
not needed for the discussion in this paper. 
The ERC-20R wrapper developer determines how these functions are implemented, their parameters, and return values. 
Similarly, the arbitration process used to decide whether a recovery request is justified is out of scope for this paper.
This topic is discussed elsewhere~\cite{ERC20r,erc792}.

\section{Designing an R-Pool} \label{sec:rpool}

We now turn to the design of an R-pool that can be used
to exchange an unsettled ERC-20R token for a settled or base version
of the same token. 
Such an exchange offloads the risk of a clawback to the 
pool's liquidity providers
who are compensated for taking on that risk. 

We explore two mechanisms for the design of such a pool.
An R-Pool can be implemented as an automated market where a decentralized risk-rating oracle assesses 
the risk of every proposed trade and provides an exchange rate on demand. 
Alternatively, the pool can be implemented in the style of an order book giving liquidity providers more control over the trades they are willing to accept. 

In the automated market implementation, a risk-rating oracle collects information from several risk-rating entities. 
The following two mechanisms help to ensure an honest and accurate rating of the risk:
\begin{itemize}[topsep=2px,itemsep=1px]
\item Every risk-rating entity must also be an LP in the pool which gives it ``skin in the game'' and incentivizes it to report the risk accurately. If it underestimates the clawback risk of the offered tokens, then the risk-rating entity harms itself in the event of a freeze or recovery. If it overestimates the clawback risk, it might lose the trade opportunity altogether.
   
\item The risk-rating entity estimates the risk of a swap based on the following information: the requestor's address and its transaction history, the ERC-20R nonce of the requestor (as explained below), and the amount to be swapped. It may also employ other off-chain sources of information. 
\end{itemize}
We next explore what swaps, recoveries, and other processes look like under both implementations.

\subsection{An automated market R-Pool}

An automated market R-Pool works similarly to an AMM, where a bonding curve contributes to determining the exchange rate. The price of a base token with respect to its ERC-20R wrapper token depends on two factors: the risk of clawback for each proposed swap, and the current liquidity of the R-Pool. Each R-Pool instance supports a particular range of risk (e.g. 0\%-20\% estimated risk of clawback). A user requests for signatures from a list of risk rating oracles qualified by the pool owner, sends them to the pool together with the swap request. Each signature contains a risk rating and address state of the user's address. The pool will only fill an order if its risk falls in the allowable risk range configured for the pool.  A liquidity provider will contribute to a pool if they are comfortable with its risk profile. 

\paragraph{Operation of the pool.} The pool only holds settled and unsettled ERC-20R tokens. When LPs add base ERC-20 tokens to the contract, the tokens immediately get wrapped as settled ERC-20R tokens. If Alice wishes to exchange her unsettled tokens for base tokens, a number of settled tokens in the pool get unwrapped and then transferred to Alice as base ERC-20 tokens. This results in a unique characteristic that does not exist in other DeFi pools: 
liquidity of the pool is self replenishing. 
The unsettled tokens that Alice sends to the pool become settled after the recoverable window elapses (say, 24 hours), thereby replenishing the pool of settled tokens that can serve as liquidity for a future user of the pool.

\paragraph{Risk rating oracles.} The risk of clawback depends on the transaction history of the address requesting the exchange,
and the amount of tokens it is trying to swap. To evaluate the risk of a swap, an off-chain aggregator service can collect risk ratings from multiple risk rating entities, which collectively form a decentralized risk rating oracle. Each risk-rating entity provides an exchange rate (e.g. between 0 and 1) for a given account requesting to exchange a specified amount of tokens. For instance, if the entity believes there is a 40\% chance of clawback, they may prescribe an exchange rate quote of 0.6.   The pool sets the exchange rate to be the median of the reported exchange rates.
We note that some existing services, such as Chainalysis, can serve as risk-rating entities. 

\paragraph{Deterring a bad actor from using the pool.}
Suppose Marvin is a bad actor who just stole some funds from a DeFi protocol, 
and the funds are sitting in Marvin's account as unsettled ERC-20R tokens.
If Marvin does nothing, then the theft will likely be clawed back by a recovery process.
Marvin, therefore, is incentivized to use the R-pool to exchange his unsettled tokens for base ERC-20 tokens.
The recovery process will then harm the pool instead of Marvin.
However, when Marvin issues his exchange request to the pool, the decentralized risk-rating oracle will recognize that the funds
recently transferred to Marvin are the result of a theft, and will rate this as a high risk exchange,
causing the pool to reject the exchange.

Marvin, however, is a clever attacker and will instead adopt the following strategy:
shortly before executing his attack, Marvin asks the risk-rating agencies to authorize an exchange from his account.
At this point, Marvin's account is pristine, and the risk rating entities will send back a high rating for the exchange.
After the high ratings are received, Marvin performs the attack transferring the stolen funds to his account.
He now uses the previously obtained high rating to exchange the unsettled tokens for settled ones using the R-pool.
This abuses the R-pool and causes the stolen funds to be clawed back from the pool. 

We prevent this attack on the R-pool using a newly introduced nonce that is associated with every ERC-20R address
(as already mentioned in \Section{sec:wrapper}).
The nonce is initialized to zero and is incremented by one every time funds are transferred to or from the associated ERC-20R address. 
Now, when a risk rating entity issues a risk report, it signs the transaction details, the risk value, and the {\em current nonce of the requestor's account}.
When the rating report is submitted on chain, the R-pool approves the transaction only if the rating is sufficiently high, 
and the nonce in the rating report is equal to the current nonce in the on chain contract.
This effectively prevents any asset transfers from taking place to or from the requestor's account for the few seconds between the time that the ratings were issued
and the time that the exchange request is posted on chain. 
If an asset transfer does take place, then the rating report becomes invalid
and the exchange request is rejected. 
This mechanism ensures that the attacker cannot execute their attack on a DeFi protocol {\em after} the rating report has been issued
for the attacker's account. 

While this nonce-based mechanism protects the R-pool, it has one negative side effect:
it prevents an honest Alice from using the R-pool as part of a long atomic transaction, say involving a flash loan.
The reason is that the flash loan deposits funds into Alice's account, thereby invalidating any previously obtained risk report for her account,
and preventing Alice from using the R-pool to exchange the flash loan tokens to base tokens.
Indeed, we claim that long atomic transactions involving flash loans enable a bad actor to abuse the R-pool,
and therefore actively prevent such swaps, even by honest parties. 

\paragraph{Liquidity incentivization and bonding curve.} 
As the settled tokens in the R-pool begin to run out, the price of base tokens becomes higher via a multiplier function.
This provides an incentive to liquidity providers to deposit more base tokens because their rewards will be higher.  
The exchange rate $r$ is thus computed, for example, by the following function:
\begin{equation} \label{eq:bonding}
  r(s,v) = R \cdot \min\left(1,\  \frac{s}{v} \cdot \frac{1}{\kappa}\right)  
\end{equation}
where $R$ is the exchange rate from the risk-rating oracle, and the remaining multiplier further reduces the exchange rate based on the state of the pool.
Here, $s$ is the number of settled ERC-20R in the pool, $v$ is the total number of ERC-20R in the pool, 
and $\kappa \in (0,1)$ is the threshold for the settled pool percentage at which the multiplier function starts to decrease from 1. 
The parameter $\kappa$ is set upon pool initialization, and can be updated by governance.
$s$ and $v$ are based on the current state of the pool. 
Another way to describe this function is that it increases linearly with respect to the settled ratio of the pool ($\frac{s}{v}$) until the satisfactory ratio threshold ($\kappa$) is reached. Other exchange rate functions can also work. 

\subsubsection{Deposit, withdrawals, and exchanges}
\label{sec:amm}
We next discuss the standard operation of an automated R-pool.

\paragraph{Deposit.}
In an automated R-Pool, liquidity providers (LPs) add base tokens to the R-Pool contract, 
and the added tokens immediately get wrapped as an ERC-20R upon entering the pool. 
In return, the LP receives LP tokens as a receipt.
Let $A>0$ be the current number of tokens locked in the pool (settled plus unsettled), 
and let $A_{\text LP}$ be the current number of outstanding LP tokens. 
If the LP deposits~$d$ base token, then $t$ LP tokens are minted and transferred to the LP, where $t$ satisfies
$t /  (A_{\text LP} + t) =  d/(A+d)$.
This ensures that the LP's fractional ownership of LP tokens (the left hand side) 
is equal to its fractional ownership of the current pool (the right hand side). 
By using the {\em current} pool size we ensure that new LPs are not penalized for clawbacks that took place before they joined the pool.
The equation for $t$ simplifies to $t \deq d \cdot (A_{\text LP}/A)$, which is well defined when $A>0$. 

\paragraph{Withdrawal.}
A liquidity provider can burn its LP tokens to withdraw its share of the liquidity pool. 
If the LP burns $t$ LP tokens then it gets back $w \deq t \cdot (A/A_{\text LP})$ tokens from the pool.
The LP receives a combination of settled and unsettled tokens that sum to $w$, 
so that the current settled-to-unsettled ratio in the pool is unchanged by the withdrawal. 
We keep the settled-to-unsettled ratio of the pool constant during withdrawal to prevent an LP from manipulating the bonding curve in Eq.~\eqref{eq:bonding}.

\paragraph{Swap.}
An ERC-20R holder will send a swap request with the amount ($P$) of unsettled tokens it intends to swap to the aggregator. Then, the aggregator will gather a quorum of $n$ risk reports and signatures from risk rating entities (the pool enforces a minimum of $n$ ratings, where $n$ is a parameter that determines the overall risk level of the pool).  Each risk-rating entity will provide a digital signature over the following tuple: address of the requestor, amount to exchange, ERC-20R nonce of the account, report expiration time, and their exchange rate quote. 
The ERC-20R holder then initiates a swap request to the automated R-Pool with the signatures and quotes. The R-Pool will check that:
\begin{itemize}[topsep=2px,itemsep=2px,parsep=2px]
    \item the number of risk reports is above the required minimum;
    \item the signer of each risk report is unique;
    \item each risk reporting entity is an LP with a minimum deposit;
    \item each risk reporting entity is authorized\footnote{The decision to allow or disallow an entity could fall on a DAO for the R-Pool.};
    \item the ERC-20R nonce of the requestor's address has not changed;
    \item none of the quotes have expired; and 
    \item the median quote is within the risk bounds of the pool.
\end{itemize} 
It then multiplies the median quote with the fee percentage based on the bonding curve in \eqref{eq:bonding} to calculate the price of the base token. 
The R-Pool then unwraps a corresponding amount ($x$) of settled ERC-20R tokens in the pool for base tokens, 
and sends $x$ base tokens to the user while receiving $P$ unsettled tokens. 
The pool size increases by $P-x$ unsettled tokens.  

\paragraph{Asset recovery.}
The risk of claw back of unsettled tokens in the pool is shared amongst the current liquidity providers, and occasionally, also by LPs who have only recently withdrawn from the pool. 
We demonstrate this using three illustrative examples. 
Consider a pool that currently holds 100 settled tokens and 100 unsettled tokens. 
A bad actor swaps out 50 settled tokens with 100 unsettled tokens, leaving the pool with 50 settled tokens and 200 unsettled tokens. 
Later, the contributed 100 unsettled tokens are frozen and clawed back through a recovery process.
Which parties incur losses as a result?
Consider the following three scenarios:
\begin{itemize}[leftmargin=5mm]
    \item After the recovery process, the pool is left with 50 settled tokens and 100 unsettled tokens. Now, a liquidity provider $L_0$ who owns 10\% of the pool withdraws all of their funds from the pool. Since the pool only has 100 unsettled and 50 settled ERC-20R tokens, the LP will receive 10 unsettled ERC-20R tokens and 5 base tokens (10\% of the unsettled and settled pool balance respectively), missing out on the 10 additional ERC-20R tokens it could have withdrawn if the pool had rejected the tainted swap.  Hence, $L_0$ incurs some losses. 
    \item A liquidity provider $L_1$ who owns 50\% of the pool withdraws all of their funds from the pool prior to the recovery process, receiving 100 unsettled ERC-20R tokens and 25 base tokens. Later, when the 100 stolen ERC-20R tokens are frozen and recovered, $L_1$ does not incur any losses because the pool has just enough ERC-20R balance to cover the 100 tokens being clawed back.  $L_1$ suffers no losses, and the LPs who have not withdrawn from the pool share the loss.
    \item A liquidity provider $L_2$ who owns 60\% of the pool withdraws all of its funds from the pool prior to the recovery process, receiving 120 unsettled ERC-20R tokens and 30 base tokens. Later, the 100 stolen ERC-20R tokens are then frozen and recovered. Since the pool only has 80 unsettled ERC-20R tokens left at the time of recovery, $L_2$ will be held accountable for the 20 tokens remaining, which means that $L_2$ still shares some of the losses despite having withdrawn from the pool before the recovery.
\end{itemize}
In summary, we see that which parties incur a loss depends on 
exact setting.

\subsubsection{The LP-shorting attack and its analysis} 

In this subsection we analyze an attack where a hacker, Marvin,
might be able to enhance its profit by interacting with an R-Pool.
We show that a well designed R-pool can prevent this from happening. 

Suppose Marvin finds a bug in a DeFi protocol~$P$.
If the protocol does not use ERC-20R wrapped tokens,
then a theft of $p$ tokens from~$P$ is unrecoverable, 
and Marvin walks away with $p$ tokens.

If protocol~$P$ uses ERC-20R wrapped tokens,
then Marvin can steal $p$ ERC-20R wrapped tokens from~$P$. 
If the R-pool risk-rating oracle does not detect the attack,
then Marvin can use the R-Pool to exchange his $p$ wrapped tokens 
for $x$ base tokens, for some $x < p$, where $x/p$ is the exchange rate offered by the R-pool.
Eventually, the attack is discovered and
protocol~$P$ initiates a recovery process to recover 
the $p$ wrapped tokens from the R-pool.
This recovery process does not affect the base token so 
in this case Marvin walks away with $x<p$ base tokens. 

So far it appears that if protocol $P$ uses wrapped ERC-20R tokens,
then Marvin's profit ($x$) is lower than his profit ($p$) if the protocol
did not use ERC-20R. 
However, when $P$ uses ERC-20R, a clever Marvin can boost 
its profits beyond $x$ by further exploiting the R-pool.
The problem is that the recovery process by protocol~$P$ 
reduces the assets of the R-pool, and
this reduces the value of the LP tokens for the R-pool.
Marvin can exploit this to increase his profits beyond $x$ by combining the 
attack on protocol~$P$ with a shorting attack on the LP tokens. 
We describe the full attack below, and then explain how to prevent it. 

\paragraph{The LP-shorting attack.}
Let $R$ be the total number of tokens, settled and unsettled, currently managed by the pool.
Let $L$ be the number of LP tokens in circulation.
We know that $R = L$ if the R-Pool has not yet experienced a recovery event;
otherwise $R < L$. 
The current value of an LP token, in base tokens, is $R/L$.
Now, consider the following sequence of actions by Marvin the hacker:
\begin{enumerate}[topsep=2px,itemsep=2px,parsep=2px]
\item 
Marvin initially puts up $c$ collateral at a lending protocol to borrow $\ell$ LP tokens. For simplicity, let's assume that $c$ is denominated in the base token used by the R-Pool; however, this could be any token accepted by the lending protocol. 

\item \label{exch1} 
Marvin sells these $\ell$ LP tokens at a DEX and receives back $b$ base tokens at the current exchange rate.  These $b$ base tokens cannot be taken
away from Marvin. 

\item 
Marvin executes his planned attack on a DeFi protocol $P$ that issued the ERC-20R wrapper. 
He steals $p$ ERC-20R wrapped tokens from~$P$. 

\item \label{exch} 
Suppose that the risk-rating oracle does not detect the attack.
Then Marvin can use the R-Pool to exchange his $p$ ERC-20R wrapped tokens for $x$ base tokens,
for some $x<p$.

\item 
Once the attack is discovered, protocol $P$ recovers $p$ wrapped
tokens from the R-pool. This decreases the value of the LP tokens because the R-Pool now holds fewer tokens. 

\item \label{exch2} 
Marvin then trades $m$ base tokens at a DEX for $\ell$ LP tokens at the new lower exchange rate. 

\item 
Marvin returns the $\ell$ LP tokens to the lending protocol and reclaims his initial $c$ collateral. 
\end{enumerate}
In this scenario, Marvin gains $x$ base tokens in step~\ref{exch}
plus $b$ tokens in step~\ref{exch1}. 
However, Marvin had to pay back his loan in $m$ base tokens
in step~\ref{exch2}. 
In total, Marvin's profit is $x + b - m$.
We are interested in when this value exceeds $p$, 
meaning that ERC-20R is worse for protocol~$P$ then ERC-20.
That is, we are interested in understanding when 
\begin{align} \label{eq:profit}
   p < x + b - m  \ .  
\end{align}
The quantities $b, \ell, m$ satisfy the following bounds:
\begin{align} \label{eq:attack}
b & = \ell \cdot \frac{R}{L} \ \ , & 
\ell & \le c \cdot \frac{L}{R} \ \ , & 
m & \ge c \cdot \frac{R-x}{R}.
\end{align} 
The middle inequality follows from the fact that $\ell/c \le L/R$,
which holds because the market rate for LP tokens $(L/R)$ is typically greater
than the borrowing rate for LP tokens $(\ell/c)$. 
The last inequality follows from the fact that after the theft,
the price of LP tokens drops by at most a factor of $\frac{R-x}{R}$
since the R-pool loses at most $x$ unsettled tokens as a result of the recovery action (see \Section{sec:amm} for a scenario where the recovery is partially covered by recently withdrawing LPs).

Using~\eqref{eq:attack}, we can bound the profitability condition~\eqref{eq:profit} as 
\[
   p < x + b - m \le x + (c - m) \le x + \frac{cx}{R} = x (1 + \frac{c}{R})
\]
which can be rewritten as 
\begin{equation} \label{eq:prof}
   \frac{x}{p} > \frac{1}{1+c/R} = \frac{R}{R + c} \geq \frac{L}{L + \ell} 
\end{equation}
The last inequality holds because $L/R \ge \ell/c$, which follows from~\eqref{eq:attack}.

Now, $\frac{x}{p} \in (0,1)$ is the exchange rate of the R-Pool, 
which is capped by the rate provided by the risk-rating oracle. 
Marvin's profits are greater than $p$ only when \eqref{eq:prof} holds.
Then, we can make the following conclusions:
\begin{itemize}
\item Since $\ell \leq L$ we know that the right hand side of \eqref{eq:prof}
cannot be less than 0.5.  
Therefore, if we require that the R-pool exchange rate ($x/p$) 
is always less than 0.5, then Marvin's profit will never exceed~$p$. 

\item If Marvin is able to short at most 10\% of all LP tokens in circulation,
then the right hand side of \eqref{eq:prof} cannot be less than  91\%.
Therefore, if we require that the  R-pool exchange rate ($x/p$) 
is always less than 91\%, then Marvin's profit will never exceed~$p$. 
\end{itemize}
Hence, by enforcing an upper bound on the R-pool exchange rate, 
we can ensure that Marvin's profit will never exceed his profit if the protocol
was not using ERC-20R wrapped tokens.
To conclude this section, we note that LP-shorting attacks also come up in the
context of re-staking, as discussed in~\cite{RestakingAttack}.

\subsection{Order book R-Pool}

A very different architecture for an R-pool is an order book.  
This design enables holders of base tokens to choose their own risk profile and price curves, rather than relying on a preset risk-rating oracle. 
This mechanism is less gas efficient (assuming matching is done on-chain~\cite{Lissy}), but it offers more flexibility and control for knowledgable liquidity providers. 
A holder of unsettled ERC-20R tokens may post an ``order'' (a minimum exchange rate they are willing to accept) to the R-Pool contract. 
Then, an LP will estimate the risk of clawback associated with this order (based on the token provenance and the requestor's transaction history) and, if they are willing to accept, 
calls the order book R-pool contract to fill the order.  This triggers a swap of unsettled for base tokens.

Unlike the automated R-Pool, there is no deposit or withdrawal action needed in the pool from LPs, because each order-book trade will pull the base tokens directly from the LP's account when the swap is filled. 
Hence, LPs do not need to lock their tokens before an order is filled.
In addition, there is no pooled risk:  an LP that fills an order assumes the entire clawback risk from the received unsettled tokens. 

\noindent
We next briefly describe the operation of the order book. 

\paragraph{User initiates an order.}
First, the user will post an initial bid by calling {\sf postBid} and providing their address, amount of unsettled tokens to swap, their minimum asking price, and expiration timestamp of their bid. This information will be stored in the contract. The user is able to cancel their bid at any point, as long as it has not yet been filled.

\paragraph{Order-matching and Swap.}
Any party that holds base tokens can listen to posted orders and analyze the bids and the associated risk to determine whether they are willing to fill the order. If so, then they can call {\sf matchBid}, specifying the order they would like to match, and their final offered quote of base tokens to sell to the user. The contract will check that the user's ERC-20R nonce has not changed, that the bid is has not expired, and that the offered quote is indeed at least the minimum asking price. Then, the contract will move unsettled tokens from the user to the liquidity provider, and move base tokens from the liquidity provider to the user. Finally, it will delete the order from the contract.

\paragraph{Recovery.}
In an order book R-Pool, the risk of claw back is assumed entirely by the liquidity provider who authorized a swap under its risk rating mechanism. For example, if a thief requests a swap of 100 ERC-20R tokens and a liquidity provider approves such a swap with a 50\% fee, the liquidity provider loses a net of 50 base tokens if the 100 ERC-20R tokens are frozen and recovered. Each LP can employ their own risk-rating strategy without affecting other LPs.

\section{Extensions} \label{sec:discussion}

\paragraph{No R-pool for non-fungible tokens (NFTs).}
The "recoverable wrapper" concept applies equally well to ERC-721 assets, where an ERC-721R wrapper protects the wrapped asset by 
allowing recovery in case of theft~\cite{ERC20r}. 
As in ERC-20R, when an asset is transferred to a new owner it remains in an unsettled state in the recipient's account during the recovery window (say, 24 hours),
and becomes settled once the window elapses.
An NFT Marketplace might refuse to accept an unsettled asset, and therefore the new owner has to wait until the asset becomes settled or it can be unwrapped.
But what if the new owner of the NFT wishes to trade it right away? 

This raises the question of an R-pool for NFTs. 
Unfortunately, we point out that an R-Pool for NFTs cannot exist because of their non-fungibility:
it is not possible to exchange an ERC-721R wrapped asset for the same version of the asset in a non-wrapped form
because there is only one version of the asset at any given time. 
Instead, we may see the emergence of marketplaces willing to trade in non-settled assets for a discount, which exists because the buyer is assuming the risk of recovery until the recovery window elapses.

\paragraph{More applications for an R-pool.}
So far we focused on end users who may want to use the R-pool to avoid delays in using a wrapped token. 
Some DeFi protocols may need to do the same and can use the R-pool in the same way.  
In addition, one can imagine new financial products motivated by R-pools.
Some risk-seeking liquidity providers may seek high-risk R-pools due to their higher APR. 
Others may offer a CDO-like product that invests in multiple R-pools at different risk levels.

\paragraph{Recoverability at alternative layers.}
Recoverability can be implemented at different layers of the blockchain stack:
\begin{itemize}[itemsep=1px,topsep=2px,parsep=1px,leftmargin=1em]
\item {\bf At the smart contract layer:} An ERC-20 smart contract can be adapted to support recoverability, as we do in the ERC-20R wrapper. 

\item {\bf At the consensus layer:} Recoverability can be implemented at the consensus layer, where a recovery process causes all validators to mark a bad transaction as canceled.
The Pilot project, discussed in Related Work (\Section{sec:related}), is an example of one such implementation. 

\item {\bf At the account layer:} Recoverability can be implemented as an account abstraction policy, where every transfer out of a protected account incurs
a 24-hour delay.
\end{itemize}

\balance
\section{Related Work} \label{sec:related}

Architectures for recoverable assets have been explored in a number of prior works. 
Eigenmann~\cite{githubreverse} drafted a contract for a reversible token that extends the ERC-20 standard. 
It uses an escrow method, where the escrow period is 30 days, 
during which the sender could recall the money at any time, with no oversight. 
This is problematic because Bob could pay Alice for a service,
and then reverse the payment 28 days later, after Alice completed the service.

The {\em Reversecoin} project~\cite{challa} from 2015 launched as a layer~1 blockchain.
It introduced a timeout period between transaction initiation and confirmation. 
Each account has an offline key pair that enables the owner to either reverse a transaction or immediately confirm it. 
This may not prevent some modern hacks: the attacker would either steal
the confirmation key, or trick the user into using the confirmation key to confirm a malicious transaction. 
The elegant Bitcoin Convenants proposal~\cite{covenants} similarly uses 
two keys (or more) to enable a vault owner to finalize or revert transactions 
from the vault 24 hours after they were posted.

More recently, \href{https://lossless.io/}{Lossless.io}~\cite{lossless}, provides an ERC-20 token wrapper,
but with a number of differences from ERC-20R.
Anyone with staked LSS tokens can monitor on-chain events for hacks, and can freeze an address if the address is involved in a hack. 
Only a single address can be frozen for a given hack, 
so one must act quickly after a hack to freeze the attacker's address before the funds are further dispersed. 
The quickest spotter is rewarded in a winner-takes-all fashion. 
The company then decides unilaterally whether the reversal is warranted,
and if so reverses the transfer. 

Finally, \href{https://pilothq.io/}{Pilot}~\cite{pilot} is an optimistic L2 built from a modified OP-stack.
The Rollup coordinator in this L2 has the ability to remove transactions from the L2 chain
using a mechanism similar to how a block is removed when a valid fraud proof is submitted. 
This enables the coordinator to undo losses caused by hacks.

\medskip
The question of swapping settled for unsettled funds comes up in a number of settings.
For example, funds received via an optimistic Rollup are subject to a seven day
dispute period, and can be viewed as unsettled funds.  Recently,
Moosavi et al.~\cite{L2} explored the question of swapping such funds for settled funds, which can be viewed as a type of R-Pool.
However, in the case of an optimistic Rollup, one can be assured that
no dispute will be submitted by verifying for themselves that all the
transactions in a Rollup block are valid.  In the case of unsettled ERC-20R tokens,
there is no way to tell for sure if the tokens will be disputed during the freeze window.

\section{Conclusions and future work} \label{sec:conclusion}

In this work we discussed the challenges in building an R-pool that can be used to exchange unsettled ERC-20R tokens for base ERC-20 tokens.
Such a pool is needed for compatibility with the rest of the DeFi ecosystem. 
The key challenge in designing an R-pool is ensuring that a bad actor cannot use the pool to shift the recovery risk to the pool's liquidity providers. 
We explored two possible designs for an R-Pool: an automated market and an order book.
The former relies on a decentralized risk-rating oracle to determine exchange rates.  It pools the risk of a clawback among the current liquidity providers. 
The latter employs an order book that gives more flexibility to liquidity providers, but also fully relegates the risk assessment to them. Either version essentially acts as a form of decentralized insurance. 

The ERC-20R architecture, as a wrapper around ERC-20 that enables asset recovery, raises many fascinating challenges in the Web3 space. 
This paper explores one aspect of this question, but many other research questions remain. 
In the future, it would be interesting to explore and compare recoverability solutions at different layers of the blockchain stack: 
at the smart contract layer, at the consensus layer, and at the account layer, as discussed in \Section{sec:discussion}. 
Another open problem is to develop mechanisms for asset recovery in a cross-chain transfer or even in a cross-rollup transfer.

\iftoggle{fullversion}{
\paragraph{\bf Acknowledgments.}
This work was partially funded by the Simons Foundation and NTT Research. We thank Dionysis Zindros for suggesting the LP-shorting attack scenario. 
}{}

\bibliographystyle{plain}
\bibliography{whitepaper}

\end{document}